\documentclass[usenatbib,useAMS]{mn2e}
\usepackage{graphics}
\usepackage{psfrag}
\usepackage{xspace}
\usepackage{amstext}
\usepackage{amsmath}
\usepackage{amssymb}
\usepackage{natbib}
\usepackage{aas_macros}
\usepackage{ulem}
\usepackage{graphicx}
\usepackage{longtable}
\usepackage{rotating}
\usepackage{float}

\newcommand{\FigDir}[1]{./#1}

\newcommand{\Msun}{\ensuremath{M_{\odot}}\xspace}

\newcommand{\Msat}{\ensuremath{M_{\rm Saturn}}\xspace}

\newcommand{\dl}{\ensuremath{D_{\rm L}}\xspace}
\newcommand{\ds}{\ensuremath{D_{\rm S}}\xspace}

\newcommand{\thE}{\ensuremath{\theta_{\rm E}}\xspace}
\newcommand{\thS}{\ensuremath{\theta_{\star}}\xspace}
\newcommand{\re}{\ensuremath{R_{E}}\xspace}
\newcommand{\umin}{\ensuremath{u_{0}}\xspace}
\newcommand{\tzero}{\ensuremath{t_{0}}\xspace}
\newcommand{\tE}{\ensuremath{t_{\rm E}}\xspace}
\newcommand{\Ml}{\ensuremath{M_{\rm L}}\xspace}
\newcommand{\Mp}{\ensuremath{M_{\rm p}}\xspace}

\newcommand{\kpc}{\ensuremath{\rm kpc}\xspace}

\newcommand{\au}{{\rm AU}\xspace}

\newcommand{\enp}{MOA-2010-BLG-353Lb\xspace}
\newcommand{\en}{MOA-2010-BLG-353\xspace}

\newlength{\voff}
\newcommand{\planetmassdash}{\ensuremath{1.03^{+1.73}_{-0.65}}\,\Msat}
\newcommand{\planetorbitdash}{\ensuremath{1.85^{+0.79}_{-0.65}}\,\au}
\newcommand{\lensmassdash}{\ensuremath{0.21^{+0.35}_{-0.13}}\,\Msun}
\newcommand{\lensdistancedash}{\ensuremath{6.21^{+1.19}_{-1.44}}\,\kpc}

\newcommand{\planetmass}{\ensuremath{0.9^{+1.6}_{-0.53}}\,\Msat}
\newcommand{\planetorbit}{\ensuremath{1.72^{+0.56}_{-0.48}}\,\au}
\newcommand{\lensmass}{\ensuremath{0.18^{+0.32}_{-0.11}}\,\Msun}
\newcommand{\lensdistance}{\ensuremath{6.43^{+1.09}_{-1.15}}\,\kpc}

\newcommand{\planetmassDsConstrained}{\ensuremath{0.78^{+1.4}_{-0.43}}\,\Msat}
\newcommand{\planetorbitDsConstrained}{\ensuremath{1.85^{+0.63}_{-0.51}}\,\au}
\newcommand{\lensmassDsConstrained}{\ensuremath{0.16^{+0.28}_{-0.09}}\,\Msun}
\newcommand{\lensdistanceDsConstrained}{\ensuremath{7.17^{+0.87}_{-1.12}}\,\kpc}

\newcommand{\Rmoa}{\ensuremath{R_{\rm MOA}}\xspace}
\newcommand{\IogleIII}{\ensuremath{I_{\rm OGLE 3}}\xspace}
\newcommand{\IogleIV}{\ensuremath{I_{\rm OGLE 4}}\xspace}
\newcommand{\VmIogleIII}{\ensuremath{(V-I)_{\rm OGLE 3}}\xspace}
\newcommand{\VmIogleIV}{\ensuremath{(V-I)_{\rm OGLE 4}}\xspace}

\begin{document}

\title[\enp: A Possible Saturn Revealed]{\enp: A Possible Saturn Revealed}

\author[Rattenbury, N.~J. et al.]
{ \parbox{0.9\linewidth}{N.~J.~Rattenbury$^{1,A}$\thanks{e-mail: n.rattenbury@auckland.ac.nz},
D.~P.~Bennett$^{2,A}$, 
T.~Sumi$^{3,A}$, 
N.~Koshimoto$^{3,A}$,  
I.~A.~Bond$^{4,A}$,  
A.~Udalski$^{5,B}$,  
F. Abe$^{6,A}$,
A. Bhattacharya$^{2,A}$,
M.~Freeman$^{7,A}$, 
A.~Fukui$^{8,A}$,  
Y.~Itow$^{6,A}$, 
M.~C.~A.~Li$^{1,A}$, 
C.~H.~Ling$^{4,A}$, 
K.~Masuda$^{6,A}$, 
Y.~Matsubara$^{6,A}$, 
Y.~Muraki$^{6,A}$, 
K.~Ohnishi$^{10,A}$, 
To.~Saito$^{11,A}$, 
A.~Sharan$^{1,A}$, 
D.~J.~Sullivan$^{12,A}$, 
D.~Suzuki$^{2,A}$, 
P.~J.~Tristram$^{13,A}$, 
S.~Koz{\l}owski$^{5,B}$,
P.~Mr{\'o}z$^{5,B}$,
P.~Pietrukowicz$^{5,B}$,
G.~Pietrzy{\'n}ski$^{5,14,B}$,
R.~Poleski$^{5,15,B}$,
D.~Skowron$^{5,B}$
J.~Skowron$^{5,B}$,
I.~Soszy{\'n}ski$^{5,B}$,
M.~K.~Szyma{\'n}ski$^{5,B}$,
K.~Ulaczyk$^{5,B}$,
{\L}.~Wyrzykowski$^{5,B}$}\\\\\\
\parbox{0.9\linewidth}{
$^1$Department of Physics, University of Auckland, Private Bag 92019, Auckland, New Zealand\\
$^{2}${Department of Physics, University of Notre Dame, Notre Dame, IN 46556, USA}\\
$^{3}${Department of Earth and Space Science, Graduate School of Science, Osaka University, 1-1 Machikaneyama, Toyonaka, Osaka 560-0043, Japan}\\
$^{4}${Institute of Natural and Mathematical Sciences, Massey University, Private Bag 102-904, North Shore Mail Centre, Auckland, New Zealand}\\
$^{5}${Warsaw University Observatory, Al. Ujazdowskie 4, 00-478 Warszawa, Poland}\\
$^{6}${Solar-Terrestrial Environment Laboratory, Nagoya University, Nagoya, 464-8601, Japan}\\
$^{7}${School of Physics, The University of New South Wales, Sydney NSW 2052, Australia}\\
$^{8}${Okayama Astrophysical Observatory, National Astronomical Observatory, 3037-5 Honjo, Kamogata, Asakuchi, Okayama 719-0232, Japan}\\
$^{9}${Department of Physics, Konan University, Nishiokamoto 8-9-1, Kobe 658-8501, Japan}\\
$^{10}${Nagano National College of Technology, Nagano 381-8550, Japan}\\
$^{11}${Tokyo Metropolitan College of Industrial Technology, Tokyo 116-8523, Japan}\\
$^{12}${School of Chemical and Physical Sciences, Victoria University, Wellington, New Zealand}\\
$^{13}${Mt. John Observatory, P.O. Box 56, Lake Tekapo 8770, New Zealand}\\
$^{14}${Universidad de Concepci\'on, Departamento de Astronomia, Casilla 160-C, Concepci\'on, Chile}\\
$^{15}${Department of Astronomy, The Ohio State University, 140 West 18th Avenue, Columbus, OH 43210,USA}\\
$^{A}${Microlensing Observations in Astrophysics (MOA) Collaboration}\\
$^{B}${Optical Gravitational Lensing Experiment (OGLE) Group}
}}
\date{Accepted ........
      Received .......;
      in original form ......}

\pubyear{2015}

\maketitle
\begin{abstract}
We report the discovery of a possible planet in microlensing event \en. This event was only recognised as having a planetary signal after the microlensing event had finished, and following a systematic analysis of all archival data for binary lens microlensing events collected to date. Data for event \en were only recorded by the high cadence observations of the OGLE and MOA survey groups. If we make the assumptions that the probability of the lens star hosting a planet of the measured mass ratio is independent of the lens star mass or distance, and that the source star is in the Galactic bulge, a probability density analysis indicates the planetary system comprises a \planetmass mass planet orbiting a \lensmass red dwarf star, \lensdistance away. The projected separation of the planet from the host star is \planetorbit. Under the additional assumption that the source is on the far side of the Galactic bulge, the probability density analysis favours a lens system comprising a slightly lighter planet.

\end{abstract}


\section{Introduction}
\label{sec:intro}
As a planet detection technique, microlensing is unique in that the peak sensitivity of microlensing to planets falls beyond the host star's snow line. The snow line is defined as the radius in the midplane of the protoplanetary disk where the temperature is below the sublimation temperature for water. Crossing the snow line, the density of solid material increases by a factor of a few \citep{2012ARA&A..50..411G}. Understanding the population of planets that form at or near the snow line is important for the core-accretion theory of planet formation \citep{2005ApJ...626.1045I}. Microlensing searches have discovered cold planets orbiting their hosts stars (see e.g. \cite{2010ApJ...710.1641S}, \cite{2012ARA&A..50..411G}) and has also provided evidence for a large population of free-floating planets -- planets without a host star or at least separated by a very large distance from a potential host \citep{2011Natur.473..349S}.

Three survey groups now routinely monitor dense stellar fields, looking for microlensing events. These groups and their operations are briefly described below.

The Microlensing Observations in Astrophysics collaboration (MOA, \cite{2001MNRAS.327..868B}, \cite{2003ApJ...591..204S}) uses the 1.8 metre MOA-II telescope at the Mount John University Observatory, Tekapo, New Zealand. The MOA-cam3 camera \citep{2008ExA....22...51S} mounted on the MOA-II telescope has a 2.2 square degree field of view and is able to observe 50 square degrees of the Galactic bulge every hour. 

The Optical Gravitational Lensing Experiment (OGLE, \cite{2015AcA....65....1U}) observe crowded stellar fields using the 1.3 metre Warsaw telescope at Las Campanas, Chile. The fourth phase of the OGLE project -- OGLE-IV -- is currently in operation and monitors over 3000 degrees of the sky. Both the MOA and OGLE survey operations include high cadence observations ($< 60$ minutes) for a subset of their fields. Such high cadence observations led to the discovery of the population of free-floating planets mentioned above and reported by \cite{2011Natur.473..349S}.

The OGLE survey detects $\sim 2000$ microlensing events every year and the MOA collaboration detects $\sim 600$ events per year. Most of the MOA detections have an OGLE counterpart and the longitudinal separation of Las Campanas and Tekapo means that these events can be well-sampled in time by these two survey groups. Both the MOA and OGLE groups issue real time alerts as new events are discovered.

The Korean Microlensing Network (KMTNet, \cite{2012SPIE.8444E..47P}) of three wide field-of-view telescopes located in South Africa, Australia and Chile has also recently started survey operations. 

During each year's microlensing ``season'' -- during the southern hemisphere winter -- microlensing event data are monitored for evidence of a deviation from that expected assuming a single mass is acting as the gravitational lens. Such deviations are found in real time by human observers, and by the expert system ARTEMiS \citep{2008AN....329..248D, 2010AN....331..671D} and the community alerted to the presence of an anomaly that may be owing to a second mass -- possibly a planet -- in the lens system. Preliminary models are circulated, including those generated automatically by the RTModel system \citep{2010MNRAS.408.2188B,2012MNRAS.424..902B}\footnote{www.fisica.unisa.it/GravitationAstrophysics/RTModel/\ldots\allowbreak{} \phantom{XXXXXXXXXXXXXXXXXXXXXX}\ldots 2014/RTModel.htm}. In response to anomaly alerts, one or more follow-up collaborations proceed to invest their observational resources to monitor events of particular interest. The follow-up groups, which include $\mu$FUN \citep{2010ApJ...720.1073G}, PLANET \citep{1998ApJ...509..687A}, RoboNet \citep{2009AN....330....4T} and MiNDSTEp \citep{2010AN....331..671D} monitor anomalous events as intensely as deemed necessary. These data, combined with the survey data, are used to constrain models of the lens system which could be found to include one or more planets. To date, 28 microlensing planets have been found by microlensing, and the reader is directed to the NASA Exoplanet Archive\footnote{exoplanetarchive.ipac.caltech.edu/index.html} for details. 

Follow-up data were obtained for most microlensing events which showed evidence of a planet or planets. In contrast, some planets have been discovered using survey data alone, e.g. events MOA-2007-BLG-197 \citep{2008ApJ...684..663B} and MOA-2008-BLG-379 \citep{2014ApJ...780..123S}. While follow-up data were collected for event MOA-2011-BLG-293, the survey data alone were sufficient to characterise the planetary system \citep{2012ApJ...755..102Y}. The planet found in event MOA-2011-BLG-322 was found using survey data from the MOA and OGLE groups and the Wise Observatory \citep{2014MNRAS.439..604S}. This present work reports another planet found using microlensing survey data alone. That the data contained a planet anomaly was only discovered after the event, once a systematic search over the binary lens parameter space was made for all archival binary lens microlensing data light curves. 

In this paper, we present the analysis of microlensing event \en. In Sections~\ref{sec:observations} and \ref{sec:reduction} we describe the data and their treatment, respectively. The results of modelling the observed light curve data with a binary lens model are presented in Section~\ref{sec:modelling}. An analysis of the background source star is given in Section~\ref{sec:sourcestar}. The planet parameters estimated from a probability density analysis is presented in Section~\ref{sec:likelihood} and Section~\ref{sec:conclusion} comprises our discussion and conclusion.

\section{Observations}

Microlensing event \en was discovered at (RA, Dec) = (18\wh05\wm12\fs94, -27\wdg17\wm35\fs64) and an alert was issued by the MOA collaboration on July 28th, 2010. The MOA collaboration observed the field in which \en was discovered with a median cadence of ~20 minutes over the course of the event. The MOA observations were made in the custom MOA-red filter, which has a passband corresponding to the sum of the standard I- and V- filters. 

During the period between May 2009 and May 2011, the OGLE survey team were not issuing alerts to the community as it was upgrading both their camera and Early Warning System (EWS) software \citep{2003AcA....53..291U}. However, the OGLE team did record I-band data for this event, with a median cadence of 60 minutes, using the OGLE-IV camera \citep{2015AcA....65....1U}. 

\label{sec:observations}

\section{Data Reduction}
\label{sec:reduction}
The images obtained by the MOA telescope were reduced using the difference imaging pipeline of \cite{2001MNRAS.327..868B}. When analysing a crowded stellar field using difference imaging, a neighbouring star, of a different colour to the target star, can affect the photometry of the target star through the differential refraction effect of the atmosphere. The degree to which the source star's photometry is affected depends on the amplitude and direction of the differential refraction. In addition to discarding extreme outliers, the MOA data were corrected for differential refraction. The final MOA light curve comprised 9130 data points. 

The OGLE data set for this event comprises 3248 I-band observations, generated from the OGLE difference imaging analysis pipeline \citep{2003AcA....53..291U}. It was not necessary to apply a correction for differential refraction to the OGLE data. 

\section{Binary Lens Modelling}
\label{sec:modelling}

An initial best-fitting binary lens model to the observational data was found from a large-scale grid search over the basic binary lens parameter space. Initial values were found for the mass ratio, $q = \Mp / \Ml $, where \Mp and \Ml are planet and host lens masses respectively,  the binary lens separation, $s$, in units of the Einstein ring radius, the source star track orientation angle $\alpha$, the minimum impact parameter \umin, the Einstein radius crossing time \tE, the epoch \tzero corresponding to the source star being positioned at \umin and the relative angular source star radius $\rho = \thS / \thE$, where \thS and \thE are the angular sizes of the source star and Einstein ring radii respectively.

Starting from the initial best-fitting binary lens solution found from the grid search, we refined the solution using a variant of the image-centred ray shooting method of \cite{2010ApJ...716.1408B}. 

We used a linear limb-darkening profile for the source star and chose limb-darkening coefficients based on an estimate of the source star's (V-I) colour. This (V-I) colour estimate is model-dependent. For this reason, we iterated the steps of finding the best-fitting binary lens model using our MCMC algorithm, deriving the corresponding (V-I) colour for the source and therefrom the source star limb-darkening coefficients, until the values of the limb-darkening coefficients ceased to change. The details of how we estimated the source colour are presented in Section~\ref{sec:sourcestar}, along with the limb-darkening coefficients used in the final model for this event.

The motion of the Earth in its orbit around the Sun can impart a parallax signal for some microlensing events \citep{1992ApJ...392..442G,1995ApJ...454L.125A}. This additional effect is generally a desirable one -- allowing a more accurate estimate of the lens star mass and distance, and thereby a more accurate estimate of the absolute values of the mass and orbital radius of planetary lens components. Parallax is more likely to be seen in microlensing events with large values of \tE. This event has $\tE \simeq $ 11 days, and is therefore {\it a priori} unlikely to to show any parallax signal. Despite this, we attempted to include microlensing parallax in our models. Not unsurprisingly, we were unable to detect any parallax signal for this event. 

\begin{figure}
\includegraphics[width=84mm]{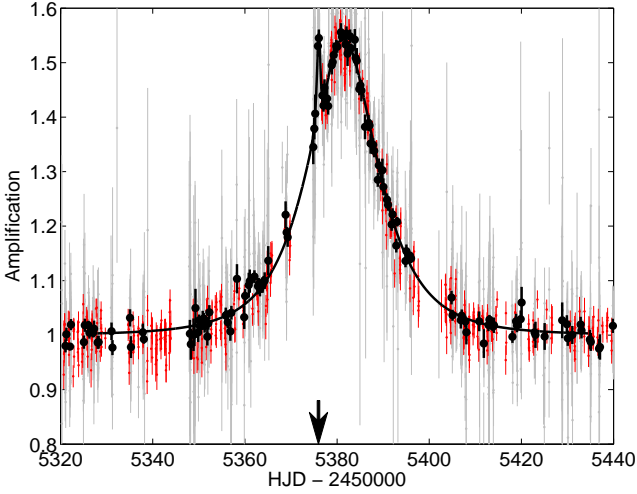}
 \caption{\label{fig:lightcurve}Observed data from the MOA (gray) and OGLE (red) microlensing survey groups for event \en. Binned MOA data points are shown as black points. The best-fitting binary lens model is also shown (black line). The parameters for this model are given in Table~\ref{tab:parameters}. The epoch when the MOA collaboration issued an alert for this event is indicated with a black arrow.}
  \end{figure}

\begin{figure}
\includegraphics[width=84mm]{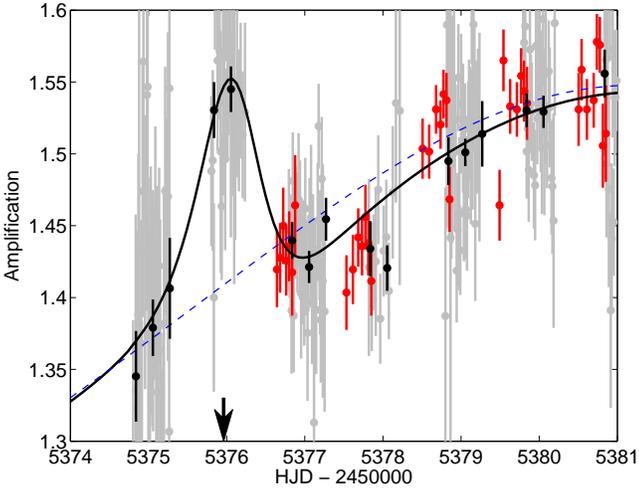}
 \caption{\label{fig:lightcurve1}This is a close-up view of the light curve for \en as shown in Figure~\ref{fig:lightcurve}, highlighting the planetary perturbation. The single lens, finite source star model is shown with a dashed blue line.}
  \end{figure}

\begin{figure}
\includegraphics[width=84mm]{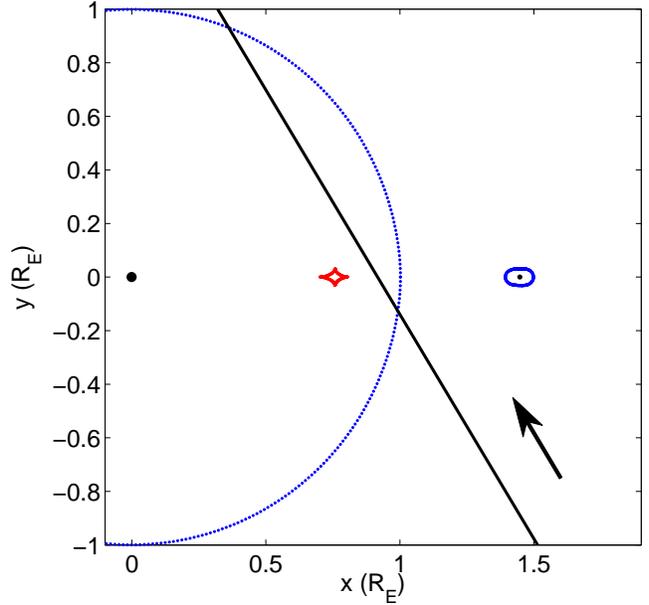}
 \caption{\label{fig:criticalcaus}The critical (blue) and caustic (red) curves corresponding to the best-fitting model for planetary microlensing event \en. The source star track is also shown (black), and the direction of the source star is indicated with an arrow.}
  \end{figure}

The best-fitting binary lens model has the following parameters for the binary lens mass ratio and binary lens separation: $q = (1.4 \pm 0.4)\times 10^{-3}$ and $s = (1.45 \pm 0.06)$. The Einstein ring radius crossing time is $\tE = (11.3 \pm 0.7)$ days. Table~\ref{tab:parameters} lists these and the other parameters in the best-fitting model. The light curve data during the microlensing event \en, along with the best-fitting light curve, is shown in Figures~\ref{fig:lightcurve} and \ref{fig:lightcurve1}. The critical and caustic curves and source star track are shown in Figure~\ref{fig:criticalcaus}.

The best-fitting planetary microlensing model is systematically above the OGLE data points at $t\simeq 5378$. To test whether this discrepancy has any significant effect on our final results, we forced the model to be consistent with the majority of these OGLE data. The parameters of this forced model are consistent with the parameters listed in Table~\ref{tab:parameters}. We ascribe the discrepancy to simple statistical fluctuations or to some unrecognised systematics.  We conclude therefore that the failure of the best-fitting model to fit the OGLE data at $t\simeq 5378$ does not significantly alter our final results.

We also investigated alternatives to the planetary microlensing model for explaining the anomaly at $t\simeq 5376$. We attempted to fit a binary source model to the data \citep{1998ApJ...506..533G} but we were unable to find a superior fit to the data. Similarly, we analysed the baseline data of this event for signs of intrinsic variability in the source. Our tests indicated that the baseline data are randomly ordered. 

\begin{table}
\caption{\label{tab:parameters}Best-fitting binary lens model parameters for planetary microlensing event \en. Each quoted error is the average of the upper and lower errors determined from the corresponding MCMC parameter state distribution.}
\begin{center}

\begin{tabular}{cc}
\hline
Parameter & Value \\
\hline
\tzero & $5381.24 \pm 0.06$ HJD - 2450000\\
\tE & $11.3 \pm 0.7$ days\\
\umin & $-0.79 \pm 0.08 \re$ \\
$q$ & $(1.4 \pm 0.4)\times 10^{-3}$\\
$s$ & $1.45 \pm 0.06$ \re\\
$\alpha$ & $4.17 \pm 0.02$ rad \\
$\rho$ & $(2.1 \pm 0.9)\times 10^{-2}$\\
\hline
\end{tabular}
\end{center}

\end{table}

\section{Source Star}
\label{sec:sourcestar}

Characterising the source star in a microlensing event leads to an estimate of the angular size of the Einstein ring in the lens plane, \thE. This in turn allows tighter constraints to be placed on the binary lens masses and lens separation, see e.g. \cite{2014ApJ...780..123S}. The value of $\rho = \thS / \thE$ is provided by the best-fitting model to the observed light curve, and an estimate of \thS may be made from known relations between optical colour and stellar radius. Determining the source star's optical colour therefore leads to our desired estimate of \thE. 

\begin{figure}
\includegraphics[width=84mm]{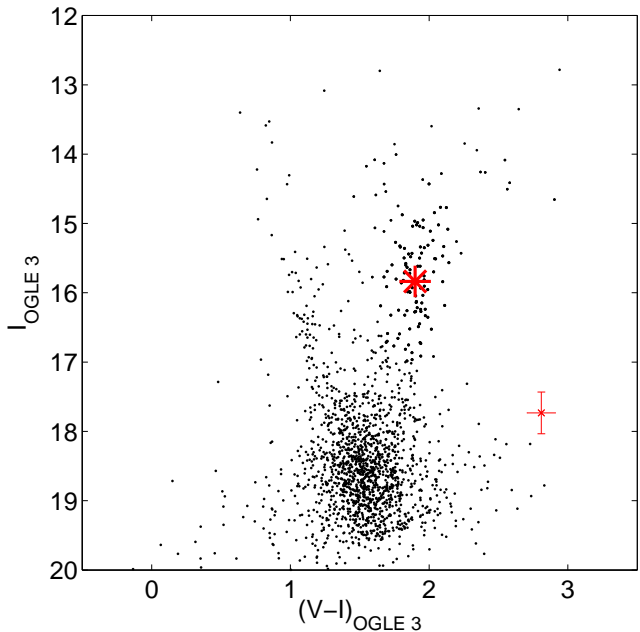}
 \caption{\label{fig:cmd}Colour magnitude diagram of the OGLE-III field stars for event \en. The centre of the red clump is indicated (asterisk) as is the location of the source star.}
  \end{figure}

The first step in determining the source star's optical colour was to cross-match MOA field stars in the 2 arcmin region surrounding the co-ordinates of event \en, with stars in the calibrated OGLE-III catalogue \citep{2008AcA....58...69U}. This led to a colour-colour relationship between the instrumental MOA and OGLE-III colours, $(\Rmoa - \IogleIII)$ and the calibrated OGLE-III colour \VmIogleIII. We similarly cross-matched MOA field stars with OGLE-IV field stars and found the relationship between $(\Rmoa - \IogleIV)$ and \VmIogleIV. Using the instrumental source colour, $(\Rmoa - \IogleIV)_{\rm S}$, obtained from the best-fitting model, we used this last relation to estimate the instrumental source colour in the OGLE-IV photometric scale, $(V-I)_{\rm OGLE 4, S}$. Cross-matching OGLE-III and OGLE-IV field stars, we fitted a straight line of the form $(I_{\rm OGLE3} - I_{\rm OGLE 4}) = a \VmIogleIV + b$. This relation, allows us to obtain $(\Rmoa - \IogleIII)_{\rm S}$ and with our first colour-colour relationship, find an estimate of the source colour $(V-I)_{\rm OGLE3, S}$ and magnitude $I_{\rm OGLE3,S}$\footnote{As the OGLE-III colours are calibrated, we will delete the OGLE-III subscript.}:
\begin{align*}
I_{\rm S} & =  17.73 \pm 0.3\\
(V-I)_{\rm S} & =  2.81 \pm 0.1 
\end{align*}
where the errors above include -- in quadrature -- errors arising from the light curve modelling, and the uncertainties arising from the colour-colour relationships. A colour-magnitude diagram of the OGLE-III field stars is shown in Figure~\ref{fig:cmd}.

We make the usual assumption that the source is located in the Galactic bulge and is therefore subject to the same reddening and extinction as bulge red clump giant stars. Red clump giant stars have tightly defined magnitudes and colours, and are therefore often used used as standard candles for Galactic structure studies \citep{2013MNRAS.434..595C}. Using a colour-magnitude diagram of the OGLE-III field stars for event \en, we found the centroid of the red clump:
\begin{align*}
I_{\rm RCG} & =  15.62 \pm 0.05\\
(V-I)_{\rm RCG} & =  1.89 \pm 0.01
\end{align*}
The intrinsic luminosity parameters of the Galactic bulge red clump are estimated in \citep{2013ApJ...769...88N}  to be
$I_{\rm RCG,0}  =  14.44 \pm 0.04$ with the colour of the red clump being centred at $(V-I)_{\rm RCG,0} =  1.06$ with dispersion $0.121$.

Comparing the observed RCG centroid colour and magnitude with these RCG intrinsic values, we find the extinction and reddening toward \en to be $(A_{\rm I}, E(V-I)) = (1.18, 0.83)\pm (0.07,0.12)$. The intrinsic source colour and magnitude are therefore $(I, V-I)_{\rm S,0} = (16.55,1.98)\pm (0.31,0.16)$. The source magnitude and colour  is consistent with a K5 subgiant in the Galactic bulge. From Table 5 of \citep{1998A&A...333..231B}, we estimate the surface temperature of a giant star with $(V-I) = 1.98$ to be $T_{\rm eff} \simeq 3750$ K. This surface temperature corresponds to linear limb-darkening coefficients in the $I$ and $R_{\rm MOA}$ bands of $(c_{\rm I},c_{R_{\rm MOA}}) = (0.5937, 0.6414)$ where $c_{R_{\rm MOA}}$ is computed as the average of the limb-darkening coefficients in the $R$ and $I$ bands. The limb-darkening coefficients differ only very slightly if we assume a subgiant source star with a cooler $T_{\rm eff} = 3500$ K.

With the intrinsic source colour in hand, we can use the relationship between optical colour and brightness-radius in \cite{2008A&A...491..855K} to estimate the source star's angular radius: $\thS = 3.97\pm 0.89$ $\mu$as. Combined with the model value of the source star angular radius relative to the angular Einstein ring radius, $\rho = \thS / \thE = (2.12 \pm 0.89)\times 10^{-2}$, we derive a value of angular Einstein ring radius, $\thE = 0.187\pm 0.089$ mas.

\section{Lens System Parameter Estimation}
\label{sec:likelihood}

Without a measurement of microlensing parallax in event \en, we have to appeal to statistical arguments in order to provide an estimate of the lens mass and distance. We followed a similar analysis to that of \cite{2006Natur.439..437B} to generate probability densities for the lens mass and lens and source distances. 

The modelled \tE and the calculated Einstein ring radius \thE are related to the lens and source distances from the observer -- \dl and \ds respectively -- and the lens mass \Ml, via the relation $\thE^2 = \kappa\Ml\left({1}/{\dl} - {1}/{\ds}\right)$ where $\kappa = 8.14$ mas$/\Msun$. We find the lens system to comprise a star with mass \lensmass with a planet of mass \planetmass. The  projected separation between the host star and the planet is \planetorbit. The distance to the lens system is \lensdistance. 

The probability density analysis was re-run, this time leaving out the constraints imposed by our estimate of \thE. The results are essentially the same as those found above. The planet has a mass of \planetmassdash and orbits a \lensmassdash star \lensdistancedash distant. The projected orbital radius is now \planetorbitdash. The planet parameters estimated from our probability density analyses are essentially independent of our characterisation of the source star. 

It is important to note that the lens system parameters found from the Bayesean probability density analysis are subject to some important assumptions. The first assumption is that the probability of the planet system comprising a planet with the measured mass ratio is independent of the mass of the host star and the host star's distance. This may not be the case in reality. We also assume that the orientation of the planetary system is random, and that the distribution of planetary radii is uniform. This affects our interpretation of the planet-star separation. If planets at large orbit radii are more common than planets with smaller orbital radii, the separation between host and planet along the line-of-sight would be much larger than the projected separation.

\subsection{A Distant Source Star}

One peculiar aspect of event \en is the colour of the source star. The very red colour of the source may, in part, be accounted for if the source star was behind the Galactic bulge, at distances greater than the usual assumed value of 8 kpc. For this reason, we re-ran the likelihood analysis, this time requiring the source star to be located on the far side of the Galactic bulge. We found the range of distances for which a G/K subgiant star would have a colour and magnitude consistent the observed source star, within the quoted errors. In estimating the range of distances for the source, we included the dust extinction profile in the direction of event \en from the high resolution 3D dust extinction map of \citet{2014A&A...566A.120S}. We found that a K5 subgiant at distances $8.25\ \kpc < \ds < 9.75\ \kpc$ has a colour and magnitude consistent with the source of \en. When we include this constraint on \ds, the lens system parameters returned by our probability density analysis -- including constraints arising from our measurement of \thE\ -- now correspond to a \planetmassDsConstrained planet orbiting a \lensmassDsConstrained star at a projected orbital radius of \planetorbitDsConstrained. The lens system is \lensdistanceDsConstrained distant.

\section{Discussion and Conclusion}
\label{sec:conclusion}
Microlensing event \en has an anomalous signal consistent with a planetary lens system. The best-fitting binary lens model for this event has a binary mass ratio of $q = (1.4 \pm 0.4)\times 10^{-3}$ and the binary lens elements are separated by $s =1.45 \pm 0.06$ \re. We were not able to find a binary source model which fits the data better than the best-fitting planetary microlensing model. The variability of the source at the baseline is consistent with random noise. 

A measurement of microlensing parallax was not possible for this event and so a well-constrained estimate of the binary lens parameters was not possible. 

While a parallax signature was not evident in the data, the finite size of the source star could be modelled. However, the nature of the source star is uncertain. The source appears much redder than the majority of field stars for this event. Given the observed source colour and magnitude, we estimate the source is a K subgiant star in the Galactic bulge. However, the source may be blended with a nearby red star, leading to an incorrect categorisation of the source star. A determination of the source star colour allowed an estimation of the angular radius of the angular Einstein ring radius. The errors on this measurement however, are large, and the \thE measurement does not serve to constrain a probability density analysis of the binary lens system parameters. 

Subject to some important assumptions, the probability density analysis for the lens system parameters suggests that the binary lens system for \en is consistent with a \planetmass planet orbiting a \lensmass M-dwarf, \lensdistance away. The projected planet-star distance is \planetorbit. One of these assumptions is that the source star resides in the Galactic bulge, at 8 kpc from our Sun. The very red colour of the source hints at the possibility that the source resides on the far side of the bulge. By constraining the position of the source star to be on the far side of the Galactic bulge, we find that our probability density analysis favours a lens system comprising a \planetmassDsConstrained planet orbiting a \lensmassDsConstrained star at a projected orbital radius of \planetorbitDsConstrained, with the lens system \lensdistanceDsConstrained distant.

The planetary signal in \en was not identified as such during the microlensing event. It was only during a subsequent systematic analysis of archived light curve data that \en was found to have an anomalous signal, possibly owing to a planet. Without a planetary anomaly alert issued in good time, no observations by the follow-up teams were made of this event. Only MOA and OGLE survey data were recorded for \en. This discovery re-emphasises the importance of conducting a systematic parameter search for all events in archival data, as pointed out by \cite{2014ApJ...780..123S}. It also demonstrates how planets can be found through high cadence microlensing survey operations alone. 

The Korean Microlensing Telescope Network epitomises this new mode of ``survey-only'' planetary microlensing. However we note that an over-reliance on solitary telescope observations, without supporting a system of contemporaneous follow-up observations, may result in missed opportunities. In \en we see the planetary anomaly essentially only in the MOA data. In the event of telescope failure, or unfavourable observing conditions, this anomaly would have gone unrecorded.

\section{Acknowledgements}
NJR is a Royal Society of New Zealand Rutherford Discovery Fellow. AS is a University of Auckland Doctoral Scholar. TS acknowledges financial support from the Japan Society for the Promotion of Science (JSPS) under grant numbers JSPS23103002, JSPS24253004 and JSPS26247023. NK is supported by Grant-in-Aid for JSPS Fellows. The MOA project is supported by JSPS grants JSPS25103508 and JSPS23340064 and by the Royal Society of New Zealand Marsden Grant MAU1104. This research has made use of the NASA Exoplanet Archive, which is operated by the California Institute of Technology, under contract with the National Aeronautics and Space Administration under the Exoplanet Exploration Program. NJR acknowledges the contribution of NeSI high-performance computing facilities to the results of this research. NZ's national facilities are provided by the NZ eScience Infrastructure and funded jointly by NeSI's collaborator institutions and through the Ministry of Business, Innovation \& Employment's Research Infrastructure programme. URL https://www.nesi.org.nz.

\bibliographystyle{latest}

\bibliography{microlensing,eclipsingBinaries,surveys,properMotion}

\end{document}